\documentstyle[11pt,newpasp,twoside,epsf]{article}
\def\edcomment#1{\iffalse\marginpar{\raggedright\sl#1\/}\else\relax\fi}
\reversemarginpar

\begin{document}
\title{HST UV and Keck HIRES Spectra of BALQSOs}
 \author{V. Junkkarinen, R. D. Cohen}
\affil{University of California, San Diego, CASS 0424, La Jolla, CA 92093-0424}
\author{T. A. Barlow}
\affil{IPAC, California Institute of Technology,
MS 100-22, Pasadena, CA 91125}
\author{F. Hamann}
\affil{University of Florida, Dept. of Astronomy, Gainesville, FL 32611-2055}

\begin{abstract}
In the analysis of broad absorption line (BAL) quasar spectra,
Keck HIRES spectra are a useful
complement to lower resolution HST and ground based spectra.
The HIRES spectra provide accurate parameters for 
narrow, intervening type absorption systems including Lyman limit
systems and a direct measurement of the smoothness of BAL features.
The smoothness of the troughs is related to the number of ``clouds''
in the BAL region if the BAL region consists of clouds.
The HIRES spectra, especially for $z$ $\sim$ 2 BAL quasars,
are also a source of high quality absorption templates.
All of the BAL features are apparently resolved in the high resolution spectra
(R $\sim$ 45 000), while low resolution (R $\sim$ 1000)
spectra do not always resolve features belonging to the outflow.
\end{abstract}

\section{Introduction}

BALs are present in about 10\% of all optically selected
quasars (Weymann et al. 1991).
The absorption features result from mass outflows with velocity widths of
order 2000 to 20 000 km s$^{-1}$ and maximum velocities to 0.1$c$ and
possibly higher in a few cases.
The chemical abundances in the BAL regions are of interest as possible indicators
of chemical enrichment in galactic nuclei (e.g. Korista et al. 1996,
Hamann and Ferland 1999).
The absorption features in BAL quasars are not easy to interpret because
the BAL region may cover only a part of the background source(s) (Hamann 1998,
Arav 1997).
Partial covering along the line of sight is common in narrower intrinsic
absorbers (e.g. Barlow, Hamann, and Sargent 1997), but more difficult to
determine in BAL quasars where the strong doublets of Si IV and C IV are
typically blended into wide features.
In narrower intrinsic absorption systems, the partial covering is both ion and velocity
dependent (Barlow and Sargent 1997).
Partial covering definitely influences some narrow BAL components as determined
from HIRES observations of CSO 755
(Barlow and Junkkarinen 1994) and Q0226$-$1024.

In this conference proceedings paper,
we report on a program that uses Keck HIRES spectra
in the analysis of the BALs.
BAL features in the UV (observed frame) are analyzed using lower resolution spectra
obtained from the HST archive.
The HIRES spectra contribute in a number of areas to the analysis.
Here we report on: 1.) BAL smoothness, and 2.) absorption
templates.
As a by--product of the template generation, the the HIRES spectra
provide some information on partial covering.

\section{Smoothness of BAL Features}

The observed smoothness of the BAL features
in HIRES spectra rules out a ``picket fence''
model for the BALs.  A ``picket fence'' model (partial covering by
turbulent components in velocity space) was proposed by Kwan (1990) in
order to explain the relative weakness of the H I Ly $\alpha$ BAL without
resorting to non--solar chemical abundances or partial covering along the line
of sight.  The smooth BAL features also imply a large number of
``clouds'' in the BAL region if the BAL region is composed of clouds
rather than a smooth outflow.

In order to quantify the smoothness of the BAL features, the C IV BAL
in Q1246$-$047 has been divided by a piecewise polynomial.
A small number of polynomials of low order were used to
preserve residual fluctuations on scales less than $\sim$ 200 km s$^{-1}$.
The result of that division is shown in Figure 1.
\begin{figure}
\plotfiddle{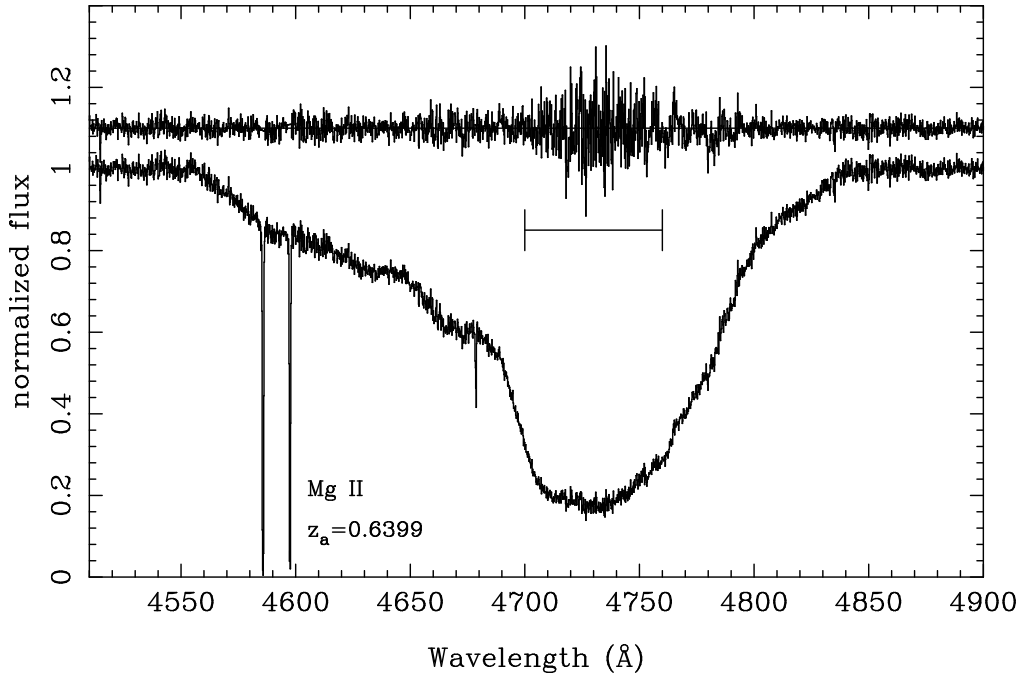}{8.3cm}{0.0}{75.0}{75.0}{-230}{-300}
\caption{HIRES spectrum, binned 4:1, of the normalized flux in the the C IV
BAL in Q1246-057.
The rms in the region indicated by the bar (4700 -- 4760\AA ) is discussed
in the text.
}
\end{figure}
The entire C IV BAL feature spans absorption velocities from 10 000 km s$^{-1}$
to 27 000 km s$^{-1}$.

In the C IV BAL, from 4700 \AA \ to 4760 \AA \ (chosen because
the optical depth is roughly constant), there are 375 8.8 km s$^{-1}$ bins.
Each bin is four pixels, about the FWHM resolution for this HIRES spectrum.
If we ignore the fact that the bins are not completely independent and
and also ignore
the doublet nature of the C IV transition, an estimate of the number of
independent ``clouds'' can be made from the noise in the divided C IV
spectrum.
The residual intensity is assumed to be given by $r$ = $e^{-\tau}$ with
$\tau$ = $N \tau_c$ where $N$ is the number of clouds contributing to the
absorption in each bin and $\tau_c$ is the optical depth per cloud.
If the clouds obey a simple gaussian distribution with its $N^{1/2}$ rms
variation in number, then the rms fluctuations
in the divided spectrum, $\sigma^2$ =  $<(\delta r / r_s)^2>$, will be
given by: $\sigma \sim \tau \delta N / N $ = $\tau N^{-1/2}$.
Here $\delta r = r - r_s$ and $r_s$ is the smooth fit to the data.
Putting in the observed $\sigma = 0.059 $ and
$\tau \sim 1$ in this wavelength range, gives
N $\sim$ 300 clouds and multiplying by 375 bins gives $N_{total} \sim 10^5$
clouds.
This is a very crude estimate, but it is clear that a large number of clouds
are needed to make a smooth trough.
The residuals observed are at a level slightly higher than expected
from the noise estimate in the spectrum (${\chi_{\nu}}^2 = 1.8$, while
1.0 is expected if purely from noise).
Given the process of piecing together the
HIRES spectral orders, fitting a continuum, and then fitting the trough with a
piecewise polynomial, it is possible that all of the observed variation
is noise plus some fitting errors.

This analysis is most sensitive to a large number of clouds
at the resolution of the observation.
Above a few hundred km s$^{-1}$, for the C IV trough in Q1246-057, 
some structure is removed by the polynomial fit used to model the smooth trough.
The bin size can easily be varied and repeating this exercise shows that
for ``cloud'' (or component) widths from 5 to 200 km s$^{-1}$ a large number
($\ge 3 \times 10^4$) of clouds is needed.
The assignment of a typical velocity
width for a typical cloud is uncertain
because the physical mechanism behind BAL cloud confinement
is not known.
Photoionization at parsec distances or greater leads to a much
larger number of very small
individual clouds and small filling factors (e.g. Junkkarinen, Burbidge,
and Smith 1987).
Velocity widths $\sim$ 8 km s$^{-1}$ (FWHM) are produced by
thermal broadening for carbon at T $\sim$ 15 000 K, near the temperature expected
from photoionization equilibrium.

The smoothness of the troughs in BAL quasars like Q1246$-$057
may instead indicate a smooth outflow.
Models that avoid the need for small clouds have been calculated by
Murray et al. (1995).
In these models the emission and absorption regions are both formed very close to
the central engine (starting point as near as $\sim$0.003 pc), 
and the absorption is produced by a smooth outflow that
crosses the line of sight at an angle.

\section{Generation of Absorption Templates}

The conventional analysis of BAL spectra involves the extraction of
absorption templates from the data.
Doublets like C IV $\lambda\lambda$ 1548.2,1550.8 and Si IV
$\lambda\lambda$ 1393.8,1402.8,
can be iteratively corrected for the weaker line
(Junkkarinen, Burbidge, and Smith 1983) to produce optical depth versus velocity.
These templates are shifted to the wavelengths of other
transitions and scaled in a $\chi^2$ fitting procedure to match the data
and give approximate (lower limit) column densities (e.g.
Korista et al. 1992).
Because BAL quasars like Q0226$-$1024 (Fig. 2) and Q1246$-$057 have smooth troughs
\begin{figure}
\plotfiddle{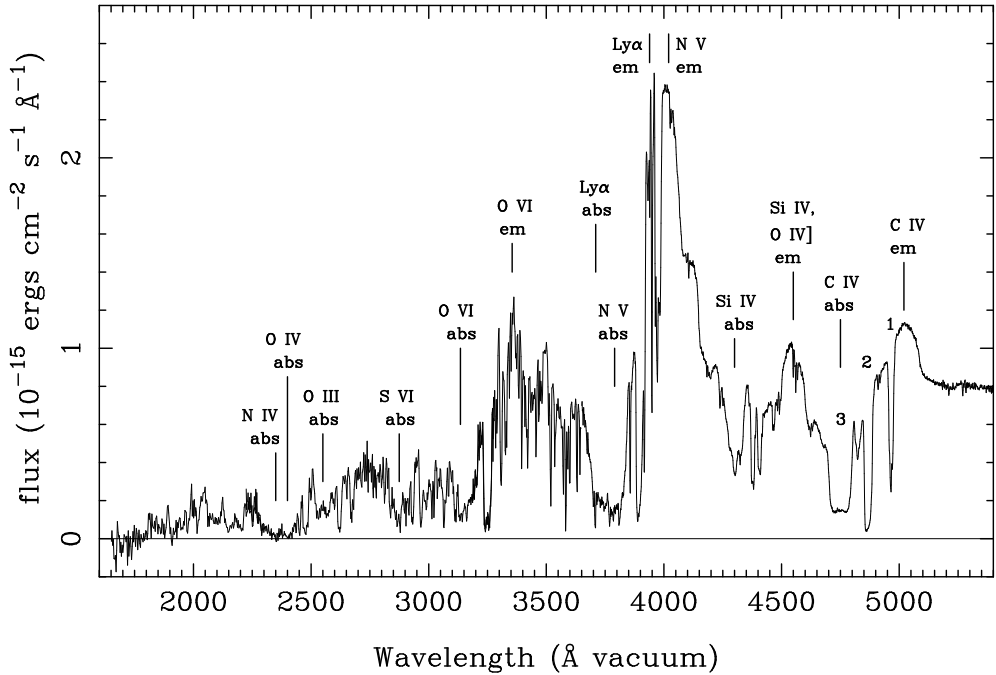}{8.7cm}{0.0}{75.0}{75.0}{-230}{-290}
\caption{Spectrum of Q0226-1024. The UV HST/FOS spectra were obtained on 1991 June 13
with the G270H grating and 1992 July 25 for the G190H grating.
The HST spectra have been smoothed using 8:1 binning for the G190H
and 4:1 binning for the G270H.
The optical spectrum was obtained on 1996 October 9 with the Lick 3m.
}
\end{figure}
with no apparent features at scales around 30 km s$^{-1}$, spectra obtained at the
HIRES resolution (typically 7 - 9 km s$^{-1}$ FWHM)
can be smoothed to produce high quality templates.

Figure 3 shows the C IV and Si IV absorption templates extracted from
\begin{figure}
\plotfiddle{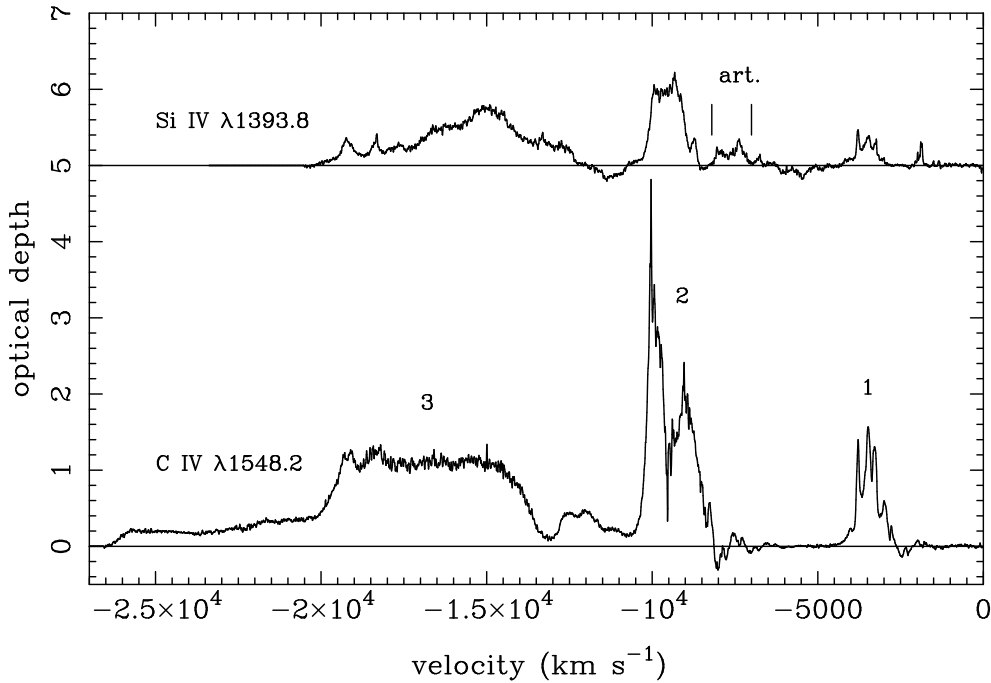}{8.4cm}{0.0}{75.0}{75.0}{-230}{-295}
\caption{Q0226-1024 absorption templates for C IV and
Si IV extracted from a Keck HIRES spectrum smoothed to 18 km s$^{-1}$ FWHM.
}
\end{figure}
our Keck HIRES spectrum of Q0226$-$1024.
The spectra were smoothed to produce these templates and could be further smoothed
without losing any of the apparent structure.
At the HIRES resolution of 7 km s$^{-1}$ FWHM for Q0226$-$1024 and at
a resolution of 9 km s$^{-1}$ FWHM for Q1246$-$057 the troughs are completely
resolved.
The troughs are not without structure as is illustrated in the case of Q0226$-$1024.
The lowest velocity BAL feature (marked ``1'' in
Figs. 2 \& 3) breaks up into three ``spikes'' of absorption repeated in
both C IV and Si IV.  These features are not resolved with $R \sim 1000$ spectra.

Analyzing BAL quasar spectra using multiple templates that cover only a part of the
velocity space may be possible when
the BALs occur in clumps well separated in velocity.
Troughs 1 and 2 in Q0226$-$1024 are narrower than the Si IV
$\lambda\lambda$ 1393.8,1402.8 doublet.
Trough 2 shows partial covering in Si IV with $C_f \approx 0.70$ based on
measurements of the doublet using:
$C_f = {{{I_1}^2 - 2 I_1 + 1} \over {I_2 - 2 I_1 + 1}}$,
where $I_1$ and $I_2$ are the residual intensities in the stronger and weaker
lines in the doublet (e.g. Hamann et al. 1997).
The template extraction procedure assumes full covering and
produces a false absorption feature around
7500 km s$^{-1}$ in the extracted template (marked ``art.''Fig. 3).
Such weak, echo--like features and regions where the template goes
negative are indications that the assumption of complete coverage of the
background source has broken down.
Artifacts can also be produced by a poor choice of continuum, so the templates
must be evaluated allowing for some uncertainty in the continuum.
For very wide features with smooth edges compared to the doublet separation, like
trough 3 in C IV in Q0226$-$1024,
it is easy to construct partial covering models that mimic the data.
The data are consistent with either 100\% covering over those velocities or with
high optical depths and a covering function
that is chosen to produce the shape of the observed feature.
The important issue of partial covering dominating the shape of the
broad BAL troughs is not one that is easily resolved.

\section{Summary}

The BALs in Q1246$-$057 and Q0226$-$1024 are
found to be smooth and completely resolved
in the Keck HIRES spectra at a resolution of 7 -- 9 km s$^{-1}$ FWHM.
The smoothness of the C IV trough in Q1246$-$057 leads to an estimate of at least
$3 \times 10^4$ clouds (or components) of width 5 to 200 km s$^{-1}$
comprising the deepest part of the trough.
Keck HIRES spectra are a good source for generating absorption templates for
analyzing lower resolution ground based and HST UV spectra of BAL quasars.
The spectrum of Q0226$-$1024 shows BAL features with structure that is not resolved in
low resolution spectra.
The Q0226$-$1024 Si IV template also shows
artifacts produced by a breakdown of the 
assumption of complete covering of the background source.
In broader BAL features (broad relative to the doublet separation), the effects of
optical depth and partial covering are not easy to separate.
Spectropolarimetry (e.g. Cohen et al. 1997) or
photoionization equilibrium analysis
could yield estimates of the partial covering and the
true column densities in the broad features.

\acknowledgments{
This work has been supported by an archival research
grant (AR-08355.01-97A) from the Space Telescope Science Institute
which is operated by AURA, Inc., under NASA contract NAS 5-26555.
}

\end{document}